# Protein contact networks at different length scales and role of hydrophobic, hydrophilic and charged residues in protein's structural organisation

**Dhriti Sengupta and Sudip Kundu***

The three dimensional structure of a protein is an outcome of the interactions of its constituent amino acids in 3D space. Considering the amino acids as nodes and the interactions among them as edges we have constructed and analyzed protein contact networks at different length scales, long and short-range. While long and short-range interactions are determined by the positions of amino acids in primary chain, the contact networks are constructed based on the 3D spatial distances of amino acids. We have further divided these networks into sub-networks of hydrophobic, hydrophilic and charged residues. Our analysis reveals that a significantly higher percentage of assortative sub-clusters of long-range hydrophobic networks helps a protein in communicating the necessary information for protein folding in one hand; on the other hand the higher values of clustering coefficients of hydrophobic sub-clusters play a major role in slowing down the process so that necessary local and global stability can be achieved through intra connectivities of the amino acid residues. Further, higher degrees of hydrophobic long-range interactions suggest their greater role in protein folding and stability. The small-range all amino acids networks have signature of hierarchy. The present analysis with other evidences suggest that in a protein's 3D conformational space, the growth of connectivity is not evolved either through preferential attachment or through random connections; rather, it follows a specific structural necessity based guiding principle – where some of the interactions are primary while the others, generated as a consequence of these primary interactions are secondary.

## Introduction

Proteins are important biomolecules having a large number of structural and functional diversities.[1] It is believed that the 3D structural and hence functional diversities of proteins, are imprinted in the primary chains of proteins. The primary chain is a linear arrangement of different amino acids connected with their nearest neighbours through peptide bonds in one dimensional space. Infact., the native 3D conformation of a protein is mainly generated and determined by the totality of inter-atomic interactions of its constituent amino-acids in 3D space. Many groups have tried to understand how primary chains of proteins consistently fold into their specific native state structures and how they attain their stabilities. Experimental studies of protein folding mechanism have been steered by several conceptual models like the framework model, diffusion-collision model, the hydrophobic collapse model and the most recent energy landscape model.[2-5] Although these different models enhance our understanding about protein folding and its structural stability, the search for a general framework or principle to explain the complex mechanism of protein folding and stability still continues.

*Department of Biophysics, Molecular Biology and Bioinformatics, University of Calcutta, 92 A.P.C. Road, Kolkata 700009, India.*
*E-mail: skbmbg@caluniv.ac.in*

In last few years, network analysis has become one of the most intriguing areas in science across many disciplines including biological systems to understand complex systems of interconnected things.[6-8] Proteins in 3D space can also be considered as complex systems emerged through the interactions of their constituent amino acids. The interactions among the amino acids within a protein can be presented as an amino acid network (often called as protein contact network) in which amino acids represent nodes and the interactions (mainly non-bonded, non-covalent) among the amino acids represent undirected edges. This representation provides a powerful framework to uncover the general organized principle of protein contact network and also to understand the sequence structure function relationship of this complex biomolecule.[9-12] Analyses of different topological parameters of protein contact networks help researchers to understand the various important aspects of a protein including its structural flexibility, key residues stabilizing its 3D structures, folding nucleus, important functional residues, mixing behavior of the amino acids and hierarchy of the structure etc.[13-19] Even, a web-server AminoNet has been recently reported to construct, visualize and calculate the topological parameters of amino acid network within a protein.[20]

The present study focuses on protein contact networks at different length scales of primary chains and the role of hydrophobic, hydrophilic and charged residues in protein folding and stability. The role of different length scales in protein folding and stability have been widely studied by several groups.[21-24] Long-range interactions are said to play a distinct role in determining the tertiary structure of a protein, as opposed to the short-range interactions, which could largely contribute to the secondary structure formations.[21-22] Taketomi and Go have concluded that specific long-range interactions are essential for highly cooperative stabilization of the native conformation while the short-range interactions accelerate the folding and unfolding transitions.[23] Sinha and Bagler have concluded that assortative mixing of long range networks may assist in speeding up of the folding process.[25] They have also observed that the average clustering coefficients of long range scales show a good negative correlation with the rate of folding, indicating that clustering of amino acids, that participate in long-range interactions, into cliques, slow down the folding process. On the other hand, several studies have been made emphasizing the dominance of hydrophobic residues in protein folding.[26-28] Poupon and Mornon have shown a striking correspondence between the conserved hydrophobic positions of a proein and the intermediates formed during its initial stages of folding constituting the folding nucleus.[29] Aftabbuddin and Kundu have also performed a comparative topological study of the hydrophobic, hydrophilic and charged residues contact network and have shown that hydrophobic residues are mostly responsible for the overall topological features of a protein.[19] Selvaraj and Gromiha have also identified the role of hydrophobic clusters in folding of (α/β)$_8$ barrel proteins and characterized the importance of medium and long-range interactions in the formation and stability of these hydrophobic clusters.[24]

When a protein folds in its native conformation, its native 3D structure is determined by the physicochemical nature of its constituent amino acids. To our knowledge, no work is reported so far to analyze the protein contact subnetworks at different length scales, which are constructed based on the physiochemical nature of amino acids and their role in protein folding and stability. These encourage our present study. Here, we have constructed and analyzed protein contact networks at two different length scales, long-range and short-range, for a large number of proteins covering all classes and folds. It should be clearly noted that while the long and short-range interactions are determined by the positions of amino acids in primary chain, the contact networks are determined by the positions of amino acids in 3D space. These long and short-range amino acids contact networks have been further divided into subnetworks of hydrophobic, hydrophilic and charged residues. Our analysis reveals a significant dominant role of hydrophobic residues over hydrophilic and charged residues in protein folding and stability. We observe that the small-range all amino acid networks exhibit a signature of heirarchy. Finally we shall discuss how the protein contact networks can be evolved in 3D space.

## Results and Discussion

We have constructed hydrophobic (BN), hydrophilic (IN), charged (CN) and all residues'(AN) networks at three different length scales; long-range interaction networks (LRNs), short-range interaction networks (SRNs) and all-range interaction networks (ARNs) for each of the 124 proteins at different interaction strength ($I_{min}$) cutoffs (see methods). We have selected subclusters having 30 or more nodes for our further analysis.[19]

**Higher degrees of hydrophobic long-range interactions suggest their greater role in protein folding and stability**

The average degree connectivities of hydrophobic, hydrophilic, charged and all residues networks of the LRNs, SRNs and ARNs for 124 proteins are calculated at different $I_{min}$ cutoffs. The values for LRN, SRN and ARN are listed in Table 1, Table 2 and Table 3, respectively.

It is evident from Table1 that the average degree connectivities ($\langle k_{LRN}^b \rangle$) of long-range BNs are higher than those ($\langle k_{LRN}^i \rangle$) of long-range INs, and are lesser than those ($\langle k_{LRN}^a \rangle$) of long-range ANs. We do not observe any long-range charged residue cluster having at least 30 nodes. For any $I_{min}$ cutoff, we observe the same trend [$\langle k_{LRN}^a \rangle > \langle k_{LRN}^b \rangle > \langle k_{LRN}^i \rangle$] (Table 1). Moreover, the sizes of the clusters of hydrophobic long-range interactions are larger than those of hydrophilic (data not shown). Both of these observations indicate that the hydrophobic interactions are playing a dominant role as compared to hydrophilic and charged interactions in long-range interaction networks. Next, we have calculated the average degree connectivities of the ANs and BNs for the SRNs (INs and CNs do not form any cluster with 30 or more node). Here, we observe that $\langle k_{SRN}^a \rangle > \langle k_{SRN}^b \rangle$ at any $I_{min}$ cutoff (Table 2). In case of ARNs, we find $\langle k_{ARN}^a \rangle > \langle k_{ARN}^b \rangle > \langle k_{ARN}^i \rangle \approx \langle k_{ARN}^c \rangle$ (average degree of charged ARNs) at any $I_{min}$ cutoff (Table 3). Further, the larger cluster sizes of ARN-BNs than ARN-INs and ARN-CNs sought for the major contribution of hydrophobic residues in protein structural organization. Similar trend has earlier been noticed by Aftabuddin and Kundu[19] in case of all-range protein contact networks, and the networks they had analyzed is equivalent to the networks studied here at $I_{min}$ =0%.

As we increase the $I_{min}$ cutoff for the different types of networks; more and more residues (nodes) in the network lose their connectivity, and as a result the average degree connectivities of the networks decrease. And at the same time, the difference between $\langle k^i \rangle$ and $\langle k^c \rangle$ values keep decreasing. However, for any $I_{min}$ cutoff, for long-range and all-range contact networks, the hydrophobic residues' connectivities are always higher than hydrophilic or charged; thus hydrophobic residues provide higher stability in a protein.

In the 3D native structure of a protein, distantly placed amino acid residues in primary chain come close to each other through long-range interactions and therefore are very important for defining the overall topology of a protein.[12,21-24] It has also been widely reported that the initiation of protein folding begins at hydrophobic sites, and that hydrophobic interactions are one of the major driving forces that folds a primary chain into its 3D structure.[26-27] Thus, the two independent set of works suggest the importance of long-range interactions and also of hydrophobic interactions in protein folding and stability. It is evident from Table 1 and Table 2 that the LRN-BNs show higher degree connectivities than the LRN-INs and SRN-BNs ($\langle k_{LRN}^b \rangle > \langle k_{LRN}^i \rangle$ and $\langle k_{LRN}^b \rangle > \langle k_{SRN}^b \rangle$). Accordingly, our result supports the leading role of LRNs and BNs in protein folding and especially reveals the dominance of hydrophobic interactions in long-range interactions which play a key role in stabilization of protein's tertiary structure. The role of long-range interactions and hydrophobic clusters are established in the folding of $(\alpha/\beta)_8$ Barrel Proteins.[24] Here we have shown the larger impact of hydrophobic residues' interactions in long-range and all amino acids' networks for a large number of proteins covering all protein classes and folds. The higher average degrees of the hydrophobic networks in ARNs and LRNs support the logic that the hydrophobic residues and the interactions among them play a major role in stabilization of protein's native conformation.

**Table 1** Number of subclusters, average degree $\langle k \rangle$, average characteristic path length $\langle L \rangle$, average clustering coefficient $\langle C \rangle$, Pearson correlation coefficient for the assortative subclusters $\langle r \rangle$, number of assortative ('pos') and disassortative ('neg') subclusters, and the ratios $p = (\langle C \rangle / \langle C_r \rangle)$ and $q = (\langle L \rangle / \langle L_r \rangle)$ of hydrophobic (BN), hydrophilic (IN), charged (CN), and all-amino-acids (AN) subnetworks in the long-range interaction networks (LRNs) are listed at different interaction strength cutoffs ($I_{min}$). No CN with atleast 30 nodes is observed.

| Cut off | Type | Number of subclusters | $\langle k \rangle$ | $\langle L \rangle$ | $\langle C \rangle$ | $\langle r \rangle$ | pos, neg | $\langle p \rangle$ | $\langle q \rangle$ |
|---|---|---|---|---|---|---|---|---|---|
| 0 | BN | 180 | 3.42 | 7.87 | 0.24 | 0.16 | 152,28 | 9.24 | 2.01 |
|   | IN | 37 | 2.48 | 6.65 | 0.12 | 0.15 | 9,28 | 2.36 | 1.54 |
|   | AN | 125 | 4.21 | 7.94 | 0.18 | 0.16 | 125,0 | 22.21 | 1.82 |
| 0.5 | BN | 180 | 3.42 | 7.87 | 0.24 | 0.16 | 152,28 | 9.24 | 2.01 |
|   | IN | 37 | 2.48 | 6.65 | 0.12 | 0.15 | 9,28 | 2.36 | 1.54 |
|   | AN | 125 | 4.21 | 7.56 | 0.18 | 0.16 | 125,0 | 22.21 | 1.74 |
| 1 | BN | 180 | 3.40 | 7.92 | 0.24 | 0.16 | 155,25 | 9.11 | 2.02 |
|   | IN | 37 | 2.48 | 6.65 | 0.12 | 0.15 | 9,28 | 2.36 | 1.54 |
|   | AN | 126 | 3.72 | 8.66 | 0.16 | 0.16 | 125,1 | 22.36 | 1.83 |
| 1.5 | BN | 206 | 3.07 | 7.53 | 0.21 | 0.13 | 147,59 | 6.43 | 1.87 |
|   | IN | 36 | 2.47 | 6.47 | 0.12 | 0.14 | 10,26 | 2.25 | 1.52 |
|   | AN | 126 | 3.49 | 9.16 | 0.16 | 0.14 | 125,1 | 22.51 | 1.85 |

**Table 2** Number of subclusters, average degree $\langle k \rangle$, average characteristic path length $\langle L \rangle$, average clustering coefficient $\langle C \rangle$, Pearson correlation coefficient for the assortative subclusters $\langle r \rangle$, number of assortative ('pos') and disassortative ('neg') subclusters, and the ratios $p = (\langle C \rangle / \langle C_r \rangle)$ and $q = (\langle L \rangle / \langle L_r \rangle)$ of hydrophobic (BN), hydrophilic (IN), charged (CN), and all-amino-acids (AN) subnetworks in the short-range interaction networks (SRNs) are listed at different interaction strength cutoffs ($I_{min}$). No IN or CN having atleast 30 nodes is observed.

| Cut offs | Type | Number of subclusters | $\langle k \rangle$ | $\langle L \rangle$ | $\langle C \rangle$ | $\langle r \rangle$ | pos, neg | $\langle p \rangle$ | $\langle q \rangle$ |
|---|---|---|---|---|---|---|---|---|---|
| 0 | BN | 35 | 2.77 | 6.76 | 0.31 | 0.12 | 9,26 | 4.05 | 1.90 |
|   | AN | 125 | 4.06 | 62.25 | 0.37 | 0.21 | 125,0 | 55.95 | 13.44 |
| 0.5 | BN | 35 | 2.77 | 6.76 | 0.31 | 0.12 | 9,26 | 4.05 | 1.90 |
|   | AN | 125 | 4.05 | 62.38 | 0.37 | 0.21 | 125,0 | 55.87 | 13.44 |
| 1 | BN | 16 | 2.64 | 6.82 | 0.27 | 0.07 | 5,11 | 3.38 | 1.84 |
|   | AN | 198 | 3.50 | 43.88 | 0.30 | 0.21 | 198,0 | 32.07 | 9.12 |
| 1.5 | BN | 12 | 2.63 | 6.68 | 0.27 | 0.09 | 3,9 | 3.54 | 1.79 |
|   | AN | 748 | 3.05 | 14.55 | 0.24 | 0.16 | 707,41 | 8.49 | 3.42 |

**Table 3** Number of subclusters, average degree $\langle k \rangle$, average characteristic path length $\langle L \rangle$, average clustering coefficient $\langle C \rangle$, Pearson correlation coefficient for the assortative subclusters $\langle r \rangle$, number of assortative ('pos') and disassortative ('neg') subclusters, and the ratios $p = (\langle C \rangle / \langle C_r \rangle)$ and $q = (\langle L \rangle / \langle L_r \rangle)$ of hydrophobic (BN), hydrophilic (IN), charged (CN), and all-amino-acids (AN) subnetworks in the all-range interaction networks (ARNs) are listed at different interaction strength cutoffs ($I_{min}$).

| Cut off | Type | Number of subclusters | $\langle k \rangle$ | $\langle L \rangle$ | $\langle C \rangle$ | $\langle r \rangle$ | pos, neg | $\langle p \rangle$ | $\langle q \rangle$ |
|---|---|---|---|---|---|---|---|---|---|
| 0 | BN | 129 | 4.95 | 7.64 | 0.39 | 0.30 | 128,1 | 18.18 | 2.25 |
|   | IN | 129 | 3.01 | 8.01 | 0.30 | 0.19 | 111,18 | 9.33 | 1.97 |
|   | CN | 77 | 2.76 | 7.15 | 0.28 | 0.21 | 59,18 | 5.20 | 1.83 |
|   | AN | 124 | 7.68 | 6.58 | 0.36 | 0.29 | 124,0 | 28.75 | 2.09 |
| 0.5 | BN | 129 | 4.95 | 7.64 | 0.39 | 0.30 | 128,1 | 18.18 | 2.25 |
|   | IN | 129 | 3.01 | 8.01 | 0.30 | 0.19 | 111,18 | 9.33 | 1.97 |
|   | CN | 77 | 2.76 | 7.15 | 0.28 | 0.21 | 59,18 | 5.20 | 1.83 |
|   | AN | 124 | 6.82 | 6.88 | 0.32 | 0.29 | 124,0 | 28.31 | 2.07 |
| 1 | BN | 136 | 4.25 | 7.90 | 0.32 | 0.24 | 133,3 | 15.96 | 2.15 |
|   | IN | 123 | 2.78 | 7.68 | 0.25 | 0.18 | 85,35 | 6.56 | 1.85 |
|   | CN | 54 | 2.64 | 6.90 | 0.25 | 0.19 | 36,18 | 4.32 | 1.75 |
|   | AN | 124 | 5.83 | 7.33 | 0.26 | 0.25 | 124,0 | 27.56 | 2.01 |
| 1.5 | BN | 142 | 3.66 | 8.45 | 0.25 | 0.19 | 135,7 | 13.33 | 2.08 |
|   | IN | 115 | 2.72 | 7.61 | 0.23 | 0.16 | 68,47 | 6.10 | 1.82 |
|   | CN | 54 | 2.64 | 6.90 | 0.25 | 0.19 | 36,18 | 4.32 | 1.75 |
|   | AN | 126 | 4.58 | 8.13 | 0.23 | 0.20 | 126,0 | 29.98 | 1.93 |

In the next sections, we intend to study and discuss how and why this dominance is important for a protein. To get a further insight view of this complex structural organization, we have calculated and compared the Pearson correlation coefficients and clustering coefficients of different protein contact subnetworks.

**Higher percentage of assortative mixing of hydrophobic residues in long-range connectivities indicates their dominant role in protein folding**

Pearson correlation coefficient ($r$) of a network is calculated to understand the mixing behaviors of its nodes. While the positive and negative $r$-values of networks suggest the assortative and disassortative mixing behaviors of the nodes, respectively;[30] it has also been reported that the information gets easily transferred through an assortative network as compared to a disassortative network.[31] Understandably, when a linear primary chain of a protein folds into its native 3D conformation, the necessary information should be communicated through the residues of that protein. Here, we shall show that the long-range hydrophobic contact networks play an important role in communicating the information.

In long-range interaction networks, the LRN-ANs show assortative mixings at lower $I_{min}$ cutoffs (Table 1). In case of LRN-BNs, 85% of the clusters show assortative mixing at lower cutoffs. The average $\langle r_{LRN}^b \rangle$ (including both positive and negative $r$ values) is 0.10, while the same calculated only for assortative networks is 0.16. Even at higher cutoffs, both LRN-BNs and LRN-ANs show high number of assortative subclusters. On the other hand, most of the INs show negative mixing behavior in LRNs with only 24% of the networks showing assortative mixing. Average $\langle r_{LRN}^i \rangle$ is -0.11, and the average $\langle r_{LRN}^i \rangle$ calculated only for assortative networks is 0.15. The LRN-CNs do not have any cluster having 30 or more nodes. In SRNs, 100% of AN clusters show assortative mixing at lower cutoffs ($\langle r_{SRN}^a \rangle \sim 0.21$), and decreases trivially at higher cutoffs (Table 2). But unlike LRNs, most of the SRN-BNs (almost 75%) show disassortative mixing of nodes.

For ARNs, our observations are similar to the results of Aftabuddin and Kundu.[19] As mentioned earlier, the networks they had analyzed is equivalent to our networks at $I_{min} = 0\%$. The present analysis have been performed for a larger set of proteins and protein contact networks at different $I_{min}$ cutoff. The ARN-ANs show assortative mixing in higher cutoffs also, while in case of ARN-BNs, more than 95% clusters show positive mixing behaviors at higher cutoffs. The ARN-INs and ARN-CNs can be assortative or disassortative, and, the numbers of assortative ARN-INs and ARN-CNs decrease at higher cutoff. Even if we consider only the positive clusters from different networks, the general trend is $r^b > r^c > r^i$.

Assortative networks are known to percolate easily, i.e. information can be easily transferred through the assortative network as compared to a disassortative network.[30] Assortative mixing tends to connect highly connected residues of a network to other residues with many contacts. Sinha et.al. have shown that the assortativities in ARNs and LRNs positively correlates to the rate of folding.[25] The ARNs and LRNs are composed of three types of subnetworks – BNs, INs and CNs. In all-range interaction networks, the BNs have the highest assortative mixing behavior indicating their major involvement in the folding process of a protein. Further, the role of long-range interactions in bringing up protein folding and stabilizing the native 3D structure is also well established.[24] As mentioned above, we find that the assortative behavior shown by the LRN-ANs is mostly contributed by its BNs (Table 1). It has been already shown in the previous section that the average degree connectivities of hydrophobics in long-range interactions are much higher than those of hydrophilics and charged. These suggest that when a protein acquires its native state, the hydrophobic residues of LRNs are the main players that pass the important information regarding folding of a protein, across the network and helps in generating the topology of protein's tertiary structure of a protein. In contrast to that, we may say that for short-range interaction networks, no specific type of residues has major contribution in communicating the necessary information; all the three types of residues (hydrophobic, hydrophilic and charged) when considered as a whole show the assortative mixing behavior, helpful in communicating the necessary information.

Thus, while short-range communication does not show preference for any specific type of residue, hydrophobics play the major role in long-range communication and thus also in tertiary structure determination or in protein folding.

**Clustering coefficients of subnetworks and their effects in protein folding and stability**

Clustering coefficient is a measure of the cliquishness of a network. The average values of clustering coefficients ($\langle C \rangle$) for long, short and all-range protein contact networks are listed in Table 1, Table 2 and Table 3, respectively. We find that ARNs and LRNs follow a similar pattern where $\langle C^b \rangle > \langle C^a \rangle > \langle C^i \rangle > \langle C^c \rangle$ at any $I_{min}$ cutoff (In LRNs, there are no charged cluster having atleast 30 or more nodes). In SRNs, average $\langle C^a \rangle$ is greater than $\langle C^b \rangle$ (Table 2). At any $I_{min}$ cutoff, we find that $\langle C_{ARN} \rangle$ is higher than $\langle C_{LRN} \rangle$ and and lower than $\langle C_{SRN} \rangle$. Sinha et al have also shown that LRNs have lower and distributed clustering coefficients than ARNs.[25]

We know that the higher value of clustering coefficient of a node '*i*' indicates the higher number of connections among it's neighbors (directly connecting nodes). Thus, the higher clustering values of nodes (amino acids) in a protein imply that the structure is more stable through the larger number of interactions among the residues. Here, we have observed that the average clustering values of hydrophobic networks are higher than those of hydrophilic and charged networks. Even we have observed that the values of $\langle C \rangle$ in LRN-BNs and ARN-BNs are higher than those of LRN-ANs and ARN-ANs, respectively. Hydrophobic residues with higher clustering values interact in a more connected fashion, stitching different secondary, super-secondary structures and stabilizing the protein structure at the global level. This further suggests the dominant role of hydrophobic residues over the others in protein stability.

It is also known that the folding of a protein and attainment of the native 3D structure is stabilized by the long-range interactions.[24] Our study shows the higher number of connectivities of hydrophobic residues in the long-range interactions those ultimately bring the distant part of the primary chain to get a specific folding and tertiary structure. Long-range interactions help in global stabilization of a protein's structure. Sinha et.al. have shown that the clustering coefficients of LRNs show a negative correlation with the rate of folding of the proteins, indicating that more time is needed for more number of mutual contacts of long-range residues for attaining the native state and hence, slower is the rate of folding.[25] As mentioned earlier, the average clustering coefficients of hydrophobic residues ($\langle C^b \rangle$) are highest in ARNs and LRNs; infact, ($\langle C^b_{LRN} \rangle$) is almost double the ($\langle C^i_{LRN} \rangle$) (no charged subcluster with required number of nodes has been observed). This also indicates that the number of times the hydrophobic residues come in loops of length three in the network is higher than the hydrophilic or charged residues, thus contributing maximum in bringing together the distant parts of a protein's linear chain. It is also very interesting to note that the values of $\langle C^a \rangle$ always lie within $\langle C^b \rangle$ and $\langle C^i \rangle$.

Based on the above observations, it is clear that the hydrophobic residues in a protein play different roles. In one hand, a significantly higher percentage of assortative subclusters of LRN-BNs and ARN-BNs (discussed in previous section) helps a protein in communicating the necessary information required for protein folding; on the other hand a higher $\langle C^b \rangle$ play a major role in slowing down the process so that necessary local and global stability can be achieved through intra connectivities of the amino acid residues.

The clustering coefficient ($\langle C \rangle$) enumerates number of loops of length three. These loops of length three can be generated by all possible combinations of hydrophobic (B), hydrophilic (I) and charged (C) residues at the vertices of a triangle. In the previous sections, we have mainly focused on BBB, III and CCC loops while studying the BNs, INs and CNs separately. Here, we have calculated and compared the number of occurrences of different triangular loops for all possible combinations of the hydrophobic, hydrophilic and charged residues at the vertices of a triangle (viz. BBB, BBI, IIC, CCC, BCI etc.) in the all-residue networks at different length scales. The number of occurrences of the loops has been normalized against the number of occurences of those residue types in the primary chain, and thus is independent of the number of B, I and C residues present in a protein. In case of LRN-ANs and ARN-ANs, about 96% of proteins show highest number of BBB loops, while the remaining 4% proteins show highest number of CCC loops. On the other hand, in SRN-ANs, maximum numbers of proteins either have highest number of CCC loops (45.96%) or have highest number of BBB loops (32.25%). This result once again supports the notion that the hydrophobic residues (rather the hydrophobic loops) play a key role in bringing the distantly placed amino acid residues along a polypeptide chain closer in the 3D space, thus shaping the overall topology of a protein. Taketomi et.al. have already concluded that specific long-range interactions are essential for highly cooperative stabilization of the native conformation.[23] This suggests the greater role of long-range hydrophobic residues in bringing up the cooperretivity and hence the stabilization of the protein three dimensional structure. It is worth mentioning that the CCC loops occur as the second highest clique of three in the LRNs and ARNs. In case of SRNs, the CCC loops occur in much higher number than the BBB loops. Thus, it is very much clear that charged loops within a protein also play a significant role in protein's structural organization.

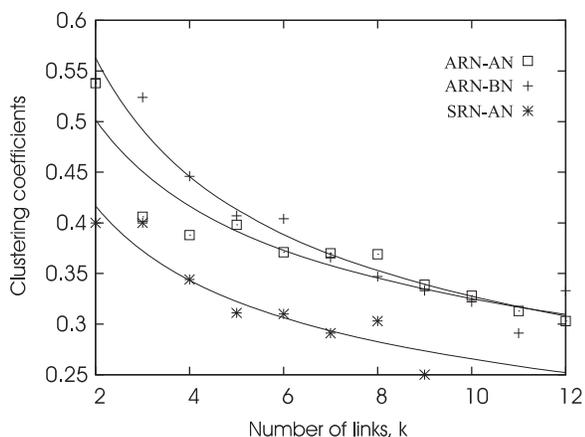

**Fig.** 1 Clustering coefficient as a function of degree 'k' for ARN-AN, ARN-BN and SRN-AN for a representative protein, 1G8K. The best-fit curves are shown by lines.

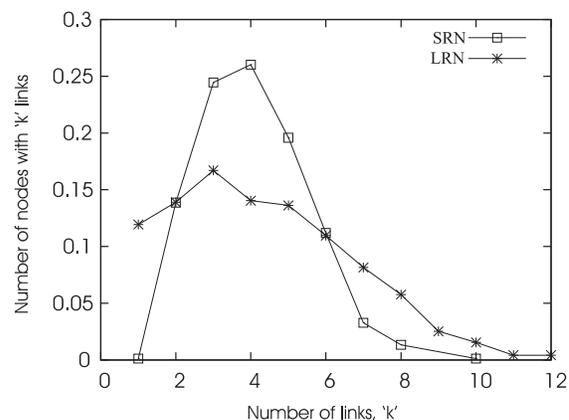

**Fig. 2** Degree distribution patterns of LRN-ANs and SRN-ANs for a representative protein, 1G8K

**ARNs and SRNs have signatures of hierarchy**

Aftabbudin anf Kundu have previously shown that ANs and BNs (in ARNs) have signatures of hierarchy.[19] Here, we shall further show that SRN-ANs also have a signature of hierarchy.

The hierarchical signature of a network lies in the scaling coefficient of $C(k) \approx k^{-\beta}$. A network is hierarchical if β has a value of 1, whereas for a nonhierarchical network the value of β is 0.[8,32] In ARN-ANs and ARN-BNs, the average values of scaling coefficient β lies neither close to 0 nor 1, but take intermediate values (varies from 0.146 to 0.356 in ARN-ANs and 0.153 to 0.600, in ARN-BNs). (Fig. 1). The values of the scaling coefficients imply that the networks have a tendency to hierarchical nature. The same observation has also been mentioned by Aftabuddin and Kundu.[19] In addition, we have observed presence of hierarchical signature in SRN-ANs, where the scaling coefficient β varies from 0.167 to 0.510 (Fig. 1). To our knowledge, we are the first to observe a hierarchical signature in short-range interaction networks.

The clustering coefficients of both the INs and CNs do not show any clear functional relation with their degree $k$ at any given $I_{min}$ cutoff. Neither of the BNs, INs and CNs generated from long-range and short-range interactions show any kind of hierarchical signature at any $I_{min}$ cutoff.

**Short-range all amino acid networks exhibit assortativity but no small world property**

We have already shown that long-range and short-range all amino acids networks have assortative mixing behavior of the nodes. Here, we shall show that while long-range and all-range different type networks have the small world properties, the short-range all amino acids network do not show any small world property.

A network is small world world if it has $C \gg C_r$ and $L \geq L_r$.[33] However, in order to show small world property, $p (= C/C_r)$ need not always be as high, there are several instances where $p$ has smaller values.[34-36] Particularly, in intra protein amino acid networks, $p$ varies from 4.61 to 25.20.[18] Similar to the observations of Aftabbudin and Kundu,[19] our results also indicate that the ARNs (ANs, BNs, INs and CNs) show small world properties at different cutoffs (Table 3). The LRN-BNs and LRN-ANs also fulfill both of the conditions of a small world network (Table 1).

On the other hand, small-range all amino acid networks having high clustering coefficients as well as high charecteristics pathlengths $(L \gg L_r)$ [Table 2] are not small world. It is expected that secondary structures of a protein (more regular networking archetecture) are generated through small-range all amino acid connectivities. However, Watts and Strogatz[33] have shown that regular networks can turn into small-world networks by the introduction of a few long-range edges. Such 'short cuts' connect vertices that are otherwise much farther apart than random networks (with smaller L). For regular networks, each short cut has a highly nonlinear effect on L, contracting the distance not only between the pair of vertices that it connects, but between their immediate neighborhoods, neighborhoods of neighborhoods and so on. Indeed, in case of a protein, when long-range interactions are added to short-range interaction networks, the resultant network (ARN) exhibits small world property.

**All-range (at higher $I_{min}$ cutoff) as well as long-range interaction networks' degree distributions are not Poisson's distribution**

We have then investigated the nature of the degree distributions of nodes for all the different type of networks at different $I_{min}$ cut-off values. At 0% $I_{min}$ cutoff, the degree distribution patterns of SRN-ANs (Fig. 2) and ARN-ANs (Fig. 3) reveal bell shaped Poisson like distributions; while for LRN-ANs (Fig. 2), we have noticed a large number of nodes with a small number of links and a small number of nodes with a large number of links. The long-range residues' connectivities pattern suggests that it cannot be generated completely through the evolving principle of random among nodes which ultimately give a Poisson's distribution. At the same time it fails to show the properties of a scale-free network.

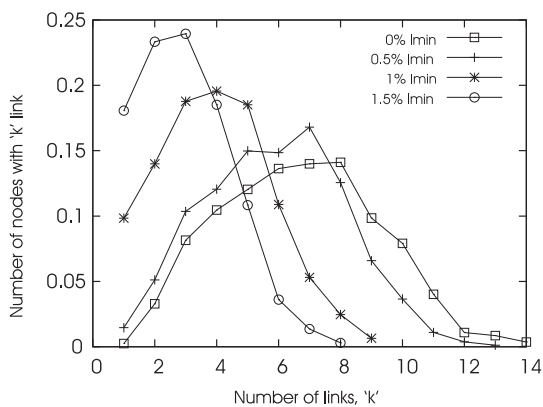

**Fig.3** The degree distribution patterns of ARNs for a representative protein 1G8K change as $I_{min}$ is increased from 0% to 1.5%.

We have further observed a similar phenomenon (as observed in Kannan and Visveswara[11]) in the degree distribution patterns of ARN-ANs at different $I_{min}$ cut-off values. The distribution pattern at 0% $I_{min}$ cutoff is Poisson like. However, when we increase $I_{min}$ cutoff the distribution patterns deviate more and more from Poisson's distribution. At the same time, these distributions at higher $I_{min}$ cutoff do not belong to the scale-free pattern (Fig. 3).

**A clue for growth of amino acids contact network in 3D space**

The information regarding the 3D structure of a protein is imprinted in the linear arrangement of its constituent amino acids in the primary chain and the said structure is evolved through interactions of amino acids in 3D space. Different proteins depending on their different compositions and arrangements of amino acids in primary chain can fold into diverse fashion. Recently, Brinda et al. have demonstrated that the observed degree distribution of protein side chain contact networks can be obtained by a principle of random connections of its constituent amino acids.[37] This indicates that an amino acid within a protein has a large degree of freedom to be connected with other amino acids. At the same time in another paper, the authors report that a random network fails to generate the number of cliques of 3 (subgraphs) as onserved in a protein contact network.[38] In this context, our observations indicate that the protein contact networks cannot be generated solely by randomly connected principle, rather suggest that the hydrophobic residues play an important role in protein folding and stability. This argument is supported by the following evidences: (i) higher average degrees of hydrophobic residues in protein contact networks, (ii) higher number of hydrophobic subclusters with assortative mixing behaviors, (iii) higher values of clustering coefficients of the hydrophobic residues, (iv) non-Poisson's' like as well as non- scale-free like distributions of the long-range connectivity networks and (v) highest occurrence of hydrophobic residues at the vertices of subgraphs of clique 3 in all and long-range interaction networks. Further, we also observe a significant role of CCC loops in protein contact networks.

Moreover, we observe that the degree distribution patterns of all-interaction networks deviate more and more from Poisson's distribution as we increase the $I_{min}$ cutoff. Understandably, at 0% $I_{min}$ cutoff an edge can exist even if there is atleast a single interaction between two amino acids. However, when we increase the $I_{min}$ cutoff, the existence of an edge needs presence of more number of non-covalent interactions. Thus, at higher $I_{min}$ cutoff, the edges with strong connections in terms of higher strength (number of possible links) which may be very crucial for the protein's structural stability and conformation are left. Interactions those are structurally so important for a protein cannot be random and accordingly, we find the degree distribution graphs shifting increasingly from the Poisson's distribution at higher $I_{min}$ cutoffs. At the same time, it should also be noted that the protein contact network's degree distribution patterns do not follow scale-free behavior either.

We have also shown here that all-amino acid's short-range interaction networks are assortative but do not follow any small world property. Small et.al. have reported a network that is highly assortative but not small world.[39] The most important scheme of that report is that the network does not grow through preferential attachment; rather it follows a specific need based guiding principle. We believe that in a protein's 3D conformational space, the growth of connectivity is evolved neither through preferential attachment nor through random connections. All connectivities cannot be generated by completely residue independent random interactions. There must be some sequence specificity. However, since all the residues are connected to their nearest neighbours through peptide bonds in a linear chain, when any residue '$i$' comes closer to another residue '$j$' in 3D space, the residues '$i$' and '$j$' force their adjacent residues in the primary chain to come closer in 3D space. Thus, one can argue that some of the interactions are primary while the others, generated as a consequence of these primary interactions are secondary; i.e., the effect of these primary interactions.

There also exists a large number of literatures supporting the preferences of different combinations of amino acid residues for different secondary structural organizations, presence of key residues for maintaining the structure and function of different proteins and also the presence of sequence motifs that are conserved in a particular family of protein.[40-41] Further it is also well established that a high fraction of coevolving amino acid residues those are important for mainiting the structural and functional integrity of proteins prefer spatial proximity in 3D space.[42] Thus, the interactions of amino acids in 3D space can not be random. We want to argue that the combined effect of the two processes - necessity driven (i.e., non-random) and random (generated through a large degree of freedom in local 3D space), could probably be the reason responsible for the Poisson like distribution in short-range interactions (it may be noted that the degree distribution is not perfectly random). On the other hand, long-range interactions are known to play important roles in determining the shape of protein tertiary structure. Most of the long-range interactions are generated through more structural necessity driven process as the said long-range interactions occur only when two distantly placed amino acids in the primary chain come close to each other in 3D space so as to stabilize the native conformation of the protein. Taketomi and Go have also shown that specific long-range interactions are essential for the highly cooperative stabilization of the native conformation.[23] The present analysis shows the higher average degree and clustering coefficients of hydrophobic residues over others in long-range interaction networks. We can thus say that the nature of connections in long-range networks are mainly non random. Moreover the higher occurrences of three sides loops of BBB followed by CCC in LRN-ANs; and CCC followed by BBB in SRN-ANs cannot be completely random or residue independent. It is also evident from our study that the necessary information for protein folding can be easily communicated within a protein mainly through the hydrophobic residues involved in long-range interactions. In all-range interaction networks, the signature of predominant roles of hydrophobic residues for protein folding information communication and protein stability are also shown here. Thus, in our opinion, the connectivities among residues are generated through necessity driven processes, but is limited by several factors- (i) once an amino acid comes closer to another amino acid, the backbone of the primary chain forces the secondary interactions (ii) interactions are possible if the two amino acids are within a specific cut-off distance and the steric hindrance limits the number of amino acids those may come closer to a specific amino acid and (iii) even different combinations of interactions are possible among amino acid residues confined within a three dimensional region.

## Conclusions

Overall, our study reveals the dominance of hydrophobic interactions in long-range interactions that plays a key role in stabilization of protein's tertiary structure. We have also observed that the LRN-BNs have a high number of assortative clusters and the assortative behaviors shown by the LRN-ANs are mostly contributed by their BNs. This suggests that when a protein acquires its native state, the highly connected hydrophobic residues of LRNs pass the important information regarding folding of protein, across the network. Higher clustering values in LRN-BNs indicate that the number of times the hydrophobic residues come in loops of length three in the network is higher than the hydrophilic or charged residues, thus contributing maximum in bringing together the distant parts of the protein linear chain. The higher clustering values of LRN-BNs also plays important role in generating the loops through their interaction and thus providing necessary stability to a protein. Short-range all amino acid networks have signature of hierarchy. They are assortative in nature, but fail to show the small world property. We havre also noticed a significant number of occurances of CCC loops indicating an important role of charged residues in proteins structural organization. Finally, we propose that the connectivities among amino acid residues in 3D space are generated by two major principle- necessity driven connections and the associated secondary connections.

## Methods

### Construction of amino acid networks

Primary structure of a protein is a linear arrangement of different types of amino acids in one-dimensional space where any amino acid is connected with its nearest neighbors through peptide bonds. But when a protein folds in its native conformation, distant amino acids in the one-dimensional chain may also come close to each other in 3D space, and hence, different non-covalent interactions are possible among them depending on their orientations in 3D space. Each protein in data set can thus be represented as a graph consisting of a set of nodes and edges, where each amino acid in the protein structure is represented as a node. These nodes (amino acids) are connected by edges based on the strength of non-covalent interactions between two amino acids.[11] The strength of interaction between two amino acid side chains is evaluated as a percentage given by:

$$INT(R_i, R_j) = \frac{n(R_i, R_j)}{\sqrt{N(R_i) \times N(R_j)}} \times 100$$

where, $n(R_i, R_j)$ is the number of distinct interacting pairs of side-chain atom between the residues $R_i$ and $R_j$, which come within a distance of 5A° (the higher cutoff for attractive London–van der Waals forces [43]) in the 3D space. $N(R_i)$ and $N(R_j)$ are the normalization factors for the residue types $R_i$ and $R_j$. The normalization factors are calculated from a set of 124 proteins, using the method described by Kannan and Vishveshwara.[10]. An important feature of such a graph is the definition of edges based on the normalized strength of interaction between the amino acid residues in proteins. The network topology of such protein structure graphs depends on the cutoff ($I_{min}$) of the interaction strength between amino acid residues used in the graph construction. Any pair of amino acid residue ($R_i$ and $R_j$) with an interaction strength of $I_{ij}$, are connected by an edge if $I_{ij} > I_{min}$ We varied $I_{min}$ from 0% to 0.5%, 1%, 1.5% etc, and protein contact networks are constructed at these different cutoffs. One should mention here that 0% cut-off is similar to the method adopted by Aftabuddin and Kundu.[19]

The data set used in this study consists of 124 protein structures obtained from the protein data bank[44] and have the following criteria:

1. Maximum percentage identity: 25.
2. Resolution: 0.0–2.0.
3. Maximum R-value: 0.2.
4. Sequence length: 500–10,000.
5. Non-x-ray entries: excluded.
6. CA-only entries: excluded.

7. CULLPDB by chain.
8. Proteins with incomplete coordinates are removed.

Crystal structures of the 124 proteins are taken for the generation of network and analysis of network properties. We have constructed the long-range residues network (LRN), short-range residues network and all-range residues network (ARN). If any amino acid $'i'$ has an interaction with any other amino acid $'j'$, whether this would be a part of the LRN or SRN depends on the distance $[x = |i - j|]$ between the $'i'-th$ and $'j'-th$ amino acids in the primary chain. If $x > 10$, LRN is produced, while if $x \leq 10$, a SRN is produced.[12,17] It is clear that $x > 0$ will provide ARN.

It is also known that each of the 20 amino acids within a protein has different side chain and different physicochemical properties. Based on it, the 20 amino acid residues are grouped into three major classes: hydrophobic (F, M, W, I, V, L, P, A), hydrophilic (N, C, Q, G, S, T, Y), and charged (R, D, E, H, K).[19] We have generated hydrophobic networks (BN) where the hydrophobic residues are considered as nodes and link between them is established if their interaction strength exceeds a particular threshold. Hydrophilic networks (IN), charged networks (CN) and all amino acid networks (AN) are constructed similarly. Our main focus is to study how the topological properties of the hydrophobic, hydrophilic and charged residues networks differ in LRNs, SRNs and ARNs. The networks thus formed have more than one subnetwork, with the number of nodes varying over a wider range. The subnetworks having at least 30 nodes have been collected and analyzed.

**Network parameters**

Each of the networks is represented as an adjacency matrix. Any element of the adjacency matrix (A), connecting the $i^{th}$ and $j^{th}$ nodes, is given as:

$a_{ij} = 1$, if $i \neq j$ and $i$ and $j$ nodes are connected by an edge:

0, if $i \neq j$ and $i$ and $j$ nodes are not connected:
0, if $i = j$

The most elementary property of a node is its degree $k$, which tells us how many links a node has with other nodes. The degree of any node '$i$' is represented by

$$k_i = \sum_j a_{ij}$$

The average degree of a network is the average of the degrees of all the nodes present in it. The spread in the number of links a node has is characterized by a distribution function $P(k)$ where $P(k) = N(k)/\sum N(k)$ where $N(k)$ is the number of nodes with $k$ links.

To observe if there is any 'small world' property in the network, one has to determine two quantities—(i) the characteristic path length $(L)$ and (ii) the clustering coefficient $(C)$. The characteristic path length $L$ of a network is the path length between two nodes averaged over all pairs of nodes. The clustering coefficient $C_i$ of a node $'i'$ is the ratio between the total number of links actually connecting its nearest neighbors whereas $k_i(k_i - 1)/2$ is the total number possible links between the nearest neighbors of node $'i'$. It is given by

$$C_i = 2e/k_i(k_i - 1)$$

In other words, clustering coefficient $C_i$ enumerates the number of loops of length three maintained by a node $i$ and its interconnected neighbors.

Clustering coefficient of the whole network is the average of all individual $C_i$'s. For a random network having $N$ number of nodes with average degree $\langle k \rangle$, the characteristic path length $(L_r)$ and the clustering coefficient $(C_r)$ have been calculated using the expressions $L_r \approx \ln N/\ln(k)$ and $C_r \approx \langle k \rangle/N$.[33] According to Watts and Strogatz,[33] if $L$ and $C$ values of a network are such that $C >> C_r$ and $L \geq L_r$, that network can be said to have the 'small world' property.

To study the tendency for nodes in networks to be connected to other nodes that are like (or unlike) them, we have calculated the Pearson correlation coefficient of the degrees at either ends of an edge. Its value has been calculated using the expression suggested by Newman[30] and is given as

$$r = \frac{M^{-1}\sum_i j_i k_i - [M^{-1}\sum_i 0.5(j_i + k_i)]^2}{M^{-1}\sum_i 0.5(j_i^2 + k_i^2) - [M^{-1}\sum_i 0.5(j_i + k_i)]^2}$$

Here $j_i$ and $k_i$ are the degrees of the vertices at the ends of the $i$ th edge, with $i = 1,.....M$. The networks having positive $r$ values are assortative in nature.

**Acknowledgement**

The authors acknowledge DIC of Calcutta University for computational facilities and UGC RFSMS for funding the research. The work is also partially supported by funds for Nanoscience and Technology, at University of Calcutta.